\documentclass[journal=jacsat,manuscript=article]{achemso}
\usepackage[version=3]{mhchem} 
\usepackage{caption}
\usepackage{float}
\usepackage{geometry}
\usepackage[numbers,super]{natbib}

\author{K\'aroly N\'emeth}
\affiliation[Illinois Institute of Technology]
{Physics Department, Illinois Institute of Technology, Chicago,  Illinois 60616, USA}
\email{nemeth@phys.iit.edu}

\author{Aditya K. Unni}
\affiliation[Illinois Institute of Technology]
{Department of Chemistry, Illinois Institute of Technology, Chicago,  Illinois 60616, USA}
\email{aunni@iit.edu}

\author{Christopher Kalnmals}
\affiliation[Illinois Institute of Technology]
{Department of Chemistry, Illinois Institute of Technology, Chicago,  Illinois 60616, USA}

\author{Carlo U. Segre}
\affiliation[Illinois Institute of Technology]
{Physics Department and CSRRI, 
Illinois Institute of Technology, Chicago,  Illinois 60616, USA}

\author{James Kaduk}
\affiliation[Illinois Institute of Technology]
{Department of Chemistry, Illinois Institute of Technology, Chicago,  Illinois 60616, USA}

\makeatletter
\DeclareRobustCommand\onlinecite{\@onlinecite}
\def\@onlinecite#1{\mciteErrorOnUnknownfalse\begingroup\let\@cite\NAT@citenum\citealp{#1}\endgroup}
\makeatother
\title{The Synthesis of Ternary Acetylides with Tellurium: Li$_{2}$TeC$_{2}$ and Na$_{2}$TeC$_{2}$}    
\date{\today}

\begin{document}

\begin{abstract}
The synthesis of ternary acetylides Li$_{2}$TeC$_{2}$ and Na$_{2}$TeC$_{2}$ is presented as the
first example of ternary acetylides with metalloid elements instead of transition metals.
The synthesis was carried out by the direct reaction of the corresponding bialkali acetylides with
tellurium powder in liquid ammonia. Alternatively, the synthesis of Na$_{2}$TeC$_{2}$ was
also carried out by the direct reaction of tellurium powder and two equivalents of 
NaC$_{2}$H in liquid ammonia leading to Na$_{2}$TeC$_{2}$ and acetylene gas through an
equilibrium containing the assumed NaTeC$_{2}$H molecules besides the reactants and the
products. 
The resulting crystalline materials were characterized by x-ray diffraction. Implications of these
new syntheses on the synthesis of other ternary acetylides with metalloid elements and transition
metals are also discussed.
\end{abstract}


\section{Introduction}

Ternary alkali metal (A) transition metal (M) acetylides with a general
formula A$_{2}$MC$_{2}$ or AMC$_{2}$ have been the subject of research in
inorganic and coordination chemistry for a few decades
\cite{ruschewitz2006ternary,buschbeck2011homoleptic,ruschewitz2003binary,brandsma1988preparative,brandsma2003best,stang2008modern}. 
Ternary acetylides synthesized so far involve AMC$_{2}$ compounds with A
$\in$ \{Li, Na, K, Rb, Cs\} and M $\in$ \{Cu, Ag, Au\} and A$_{2}$MC$_{2}$
compounds with A $\in$ \{Na, K, Rb, Cs\} and M $\in$ \{Pt, Pd\}.  A major
source of synthesis methods and characterization results on ternary
acetylides is from the Ruschewitz
group\cite{ruschewitz2006ternary,ruschewitz2003binary,kockelmann1999novel,offermanns2000syntheses,ruschewitz2001ternare,billetter2010ternary,hemmersbach2001ternary,hamberger2012evidence}. 
It was theoretically predicted in a recent publication by Terdik,
N{\'e}meth, Harkay {\it et al} \cite{terdik2012anomalous} that the
photoemissive properties of Cs$_{2}$Te, an old photoemissive material with
very high quantum efficiency, could be improved by acetylation leading to a
predicted Cs$_{2}$TeC$_{2}$ material.  Related compounds, such
as Na$_{2}$TeC$_{2}$, might have similarly advantageous photoemissive
properties, such as high quantum-efficiency and reduced workfunction as
calculated for the parent alkali tellurides.  Existing A$_{2}$MC$_{2}$
ternary acetylides are built of linear polymeric chains with -M-C$\equiv$C- repeating
units embedded in a matrix of alkali cations.  The existence of such
chains with -Te-C$\equiv$C- repeating units was proposed by the
above-mentioned theoretical study, motivated by knowledge of A$_{2}$MC$_{2}$
compounds.  Analogous monomeric acetylenic tellurides/selenides/sulfides with the
R-C$\equiv$C-(Te/Se/S)-C$\equiv$C-R$^{\prime}$ structure have been known for
decades\cite{radchenko1989acetylenic,gedridge1992preparation,brandsma1988preparative}. 
Note, however, that in A$_{2}$MC$_{2}$ crystals the M atoms are formally neutral, and the -M-C$\equiv$C-
repeating units carry a charge of
negative two that is charge-balanced by two adjacent alkali cations, as
compared to the charge neutral R-C$\equiv$C-(Te/Se/S)-C$\equiv$C-R$^{\prime}$ species where the formal
oxidation state of the chalcogen is +2.  Alternatively, 
in the A$_{2}$TeC$_{2}$ systems developed in this study,
the -Te-C$\equiv$C- repeating unit must carry a -2 charge. While
binary transition metal acetylides with the formula MC$_{2}$ are well known
explosives, their alkalinated versions AMC$_{2}$ and A$_{2}$MC$_{2}$ are not
explosive, however, they are very air and moisture sensitive as they are
strong reducing agents \cite{ruschewitz2006ternary}.  In fact, it is because of their strong reducing
nature that LiAuC$_{2}$ and LiAgC$_{2}$,
the only known Li-containing ternary acetylides \cite{kockelmann1999novel,offermanns2000syntheses}
have even been proposed as anode materials for Li-ion
batteries \cite{pottgen2010lithium}.  Developing new syntheses of ternary acetylides with lighter and
abundant transition metals or metalloids would allow for their practical consideration as anode
materials in Li/Na-ion batteries. The goal of the present work was to develop an alternative synthesis
to these intriguing materials, and determine the structure of a new class of tellurium containing
ternary acetylides.

\section{Experimental Section}   
\subsection{Synthesis of Li$_{2}$TeC$_{2}$ from Li$_{2}$C$_{2}$ and Te}

Li$_{2}$C$_{2}$ was synthesized by the reaction of Li metal and acetylene
gas in liquid ammonia (lq-NH$_{3}$), followed by spontaneous
disproportionation of LiC$_{2}$H to Li$_{2}$C$_{2}$ and acetylene gas upon evaporation of the ammonia,
and warming to ambient temperature, as
described e.g.  in Ref.  \onlinecite{ruschewitz2003binary}.  In a N$_{2}$  glovebox, tellurium powder
(384 mg, 3 mmol) as obtained from Sigma-Aldrich (99.8 \%) was mixed with Li$_{2}$C$_{2}$ powder (120 mg,
3.17 mmol, 5.5 \% excess) in a flame-dried 100 ml Schlenk flask with a Teflon plug valve, then sealed
and brought into a fume hood.  NH$_{3}$ gas from a tank was
condensed at $\approx$ -60$^{\rm o}$C until it fully covered the solid reactants.  The chilling of the
flask was carried out in an acetone bath placed in a sonicator bay (Branson
Model 3510 Ultrasonic Cleaner) through a coil attached to an electric
laboratory chiller.  Sonication of the reactants was started when they were
completely covered by lq-NH$_{3}$ (5-10 mL).  The sonication and condensation of
NH$_{3}$ (at a slightly higher than atmospheric pressure) continued until
all the Te powder was consumed by the reaction.  The flask was occasionally
swirled by hand and lifted out of the bath to confirm the disappearance of the Te powder at the bottom
of the flask.  Depending on the concentration of the
reactants in the lq-NH$_{3}$, complete consumption of the Te typically took 10-30 minutes.  The color of
the reaction mixture gradually changed from purple to violet and finally to yellow.  When the Te was
fully consumed, the flask was gradually warmed to room temperature and the NH$_{3}$ carefully
evaporated.  The flask was then sealed under a NH$_{3}$ atmosphere and brought into the glove box for
handling and storage.

\subsection{Synthesis of Na$_{2}$TeC$_{2}$ from Na$_{2}$C$_{2}$ and Te}

Na$_{2}$C$_{2}$ was synthesized by the reaction of elemental sodium and
acetylene gas in lq-NH$_{3}$ followed by heating the resulting NaC$_{2}$H at
155 $^{\rm o}$C under vacuum as described in Ref. 
\onlinecite{kloss2002neue}.  The reaction of Te powder (384 mg, 3 mmol) with
Na$_{2}$C$_{2}$ powder (220 mg, 3.14 mmol, 5 \% excess) in lq-NH$_{3}$ was
carried out analogously to that of the above described synthesis of
Li$_{2}$TeC$_{2}$.  The resulting product is a yellow powder that also has
to be handled in an inert atmosphere.

\subsection{Synthesis of Na$_{2}$TeC$_{2}$ from NaC$_{2}$H and Te}

NaC$_{2}$H was synthesized by the direct reaction of Na (200 mg, 8.70 mmol)
with acetylene gas in lq-NH$_{3}$ using a 100 ml Schlenk flask as above. 
Tellurium powder (520 mg, 4.06 mmol) was added to the dry NaC$_{2}$H under
protective flow of NH$_{3}$ gas and lq-NH$_{3}$ was condensed again on the
reactants.  The flask was occasionally manually swirled to allow for
better mixing of the reactants.  The Te powder was consumed in about 10
minutes without sonication producing a yellowish solution with some
yellowish precipitates.  The NH$_{3}$ solvent was carefully evaporated and a
pale yellow powder obtained.  The sealed Schlenk flask was brought in the
glove box and its contents were scraped into a mortar and ground whereby
they became a yellow powder.  Powder x-ray diffraction confirms that the
materials obtained by the two different syntheses of Na$_{2}$TeC$_{2}$ are
identical.

\subsection{Synthesis of Tellurium Acetylide, TeC$_{2}$}

We have also attempted to synthesize the binary acetylide TeC$_{2}$ polymer
by the polycondensation reaction of diiodoacetylene
\cite{vaughn1932iodination} and lithium telluride \cite{smith2000syntheses}
in dry tetrahydrofuran, according to \begin{equation} \label{TeC2} n
I-C{\equiv}C-I + n Li_{2}Te \rightarrow (-Te-C{\equiv}C-)_{n} + 2n LiI . 
\end{equation} The reaction product was a metallic gray powder that was
extremely sensitive to mechanical agitation after being washed with water
and diethyl ether on filter paper.  When dry, this material violently
decomposed upon contact with a spatula.  Because of its explositivity that
is similar to that observed in transition metal acetylides, this product
could not be characterized.

\subsection{Synthesis of Ph-C$\equiv$C-Te-C$\equiv$C-Ph and its double lithiated form}

Bis(phenylethynyl)telluride, Ph-C$\equiv$C-Te-C$\equiv$C-Ph, was
synthesized following the procedure of Ref.  \onlinecite{caracelli20104},
starting from phenylacetylene (Sigma Aldrich).  We have also attempted to
synthesize its double lithiated form, in analogy to the A$_{2}$TeC$_{2}$
ternary acetylides, reacting Li-Ph-acetylide (Ph-C$\equiv$C$^{-}$Li$^{+}$)
with Te powder in a 2:1 molar ratio in a sonicated dry THF solution.  After
about 24 hours of sonication all the Te powder was consumed.  Evaporation of
THF resulted in a yellow powder which turned out to be a 1:1 molar mixture
of Ph-C$\equiv$C-Te$^{-}$Li$^{+}$ and unreacted Ph-C$\equiv$C$^{-}$Li$^{+}$,
as characterized by Extended X-ray Absorption Fine Structure (EXAFS)
spectroscopy.  This indicates that the hypothesized salt
Li$_{2}$[Ph-C$\equiv$C-Te-C$\equiv$C-Ph] does not form, and the negatively
charged -C$\equiv$C-Te-C$\equiv$C- structural units are stable only in the
A$_{2}$TeC$_{2}$ ternary acetylides.

\subsection{X-ray diffraction based characterization}

Powder x-ray diffraction (XRD) spectra of the materials produced were
obtained on a Bruker D2 Phaser benchtop XRD system using Cu-K$_{\alpha 1}$
radiation at 1.5406 {\AA} wavelength, at room temperature.  An air-tight
PMMA specimen holder with a dome (Bruker A100B33) was used to hold the
air-sensitive materials during the measurements.  This sample holder adds a
large background to the measured spectra between 15 and 25 ($^{\rm o}$)
2$\Theta$ values.  
Note that XRD data for 2$\Theta$ values smaller than those shown in Figs. \ref{XRD-Li2TeC2}
and \ref{XRD-Na2TeC2} are all due to the sample holder, including a sharper
peak at about 13 and a broader one at about 10 ($^{\rm o}$) 2$\Theta$. 
Crystal structures were refined using the GSAS software package.\cite{GSAS}

\subsection{EXAFS based characterization}

Extended X-ray Absorption Fine Structure (EXAFS) spectroscopy measurements
were carried out on Ph-C$\equiv$C-Te-C$\equiv$C-Ph and its double lithiated
form at the Te K$_{\alpha}$ edge using synchrotron radiation at the MRCAT
(Sector 10) beamline of the APS facility at Argonne National Laboratory. 
Data was processed using the IFEFFIT-based programs Athena and 
Artemis.\cite{newville2001,ravel2005}

\section{Results and Discussion}   

All previously known ternary acetylides of the A$_{2}$MC$_{2}$ type were
synthesized by the solid state reaction of the corresponding bialkali
acetylides and Pt or Pd sponges
\cite{hemmersbach2001ternary,ruschewitz2001ternare,billetter2010ternary}:
\begin{equation} A_{2}C_{2} + M \rightarrow A_{2}MC_{2} .  \end{equation} As
our initial attempts to follow the solid state reaction method failed for
the direct reaction of tellurium powder with Li$_{2}$C$_{2}$, we had been
looking for a suitable solvent to carry out the same reaction.

A recent article by the Ruschewitz group pointed out that (bi)alkali
acetylides actually dissolve to some extent in liquid ammonia (lq-NH$_{3}$)
\cite{hamberger2012evidence}.  The same study also suggested that
lq-NH$_{3}$ might be a suitable reaction medium for the synthesis of ternary
acetylides.  Motivated by this publication, we attempted the direct reaction
of Te powder with slight molar excess of Li$_{2}$C$_{2}$ or Na$_{2}$C$_{2}$
in lq-NH$_{3}$.  The rather quick dissolution of the Te powder during the reaction indicated the
reaction of the dialkali acetylides with the Te metal, as no dissolution occurs in the absence of the
acetylides. 
When run in an ultrasonicator, the Te was completely consumed in 10-30 minutes, depending on the
concentration of the reactants.  Ultrasonication significantly shortens the time needed for completing
the consumption of Te powder by the bialkali acetylide, as compared to simple stirring (2-3 hours).  We
have used various temperatures from -80 to -35 $^{\rm o}$C and the reaction always completed very
quickly when sonication was used. We have also tried using dry tetrahydrofuran and diethyl ether as a
reaction solvent but these proved far inferior to lq-NH$_{3}$ and resulted only in some discoloration of
the reactants even after hours of sonication.

The synthesis starts with the condensation of ammonia onto a mix of the Te
and bialkali acetylide powder in a suitable, dry reaction vessel, such as a
Schlenk flask.  In the initial phase of the condensation, when there is only
a small amount of lq-NH$_{3}$ present, purple/violet/red colors can be
observed in the reaction mixture.  These colors are typical indicators of
polytelluride cluster anions that dissolve well in lq-NH$_{3}$, as known for
a century \cite{smith2000syntheses}.  Such polytelluride anions may consist
of tens of Te atoms in a single cluster carrying typically only a negative
charge or two \cite{smith2000syntheses}.  In the case of our reactions, the
charge of these Te clusters may only come from acetylide ions that attack
the Te powder from the solution and split it up into cluster anions with
attached acetylide units that dissolve in lq-NH$_{3}$.  As the dissolution
of binary acetylides in lq-NH$_{3}$ is very limited
\cite{hamberger2012evidence}, one might expect a slow reaction.  However,
the small amount of dissolved acetylides is sufficient to solubilize all the
tellurium powder and then it will be the dissolved polytelluride anions that
attack the solid crystallites of bialkali acetylides to further split the
polytelluride anions until all acetylide ions are attached to Te atoms or
clusters.  Thus, the solubility of polytelluride ions allows for an
autocatalytic process and the reaction goes quickly even at low temperature,
especially when sonication assists the mixing of the reactants and their
attacks on each other.

The final product is a colloid of yellow Li$_{2}$TeC$_{2}$ or
Na$_{2}$TeC$_{2}$ crystallites in lq-NH$_{3}$, which must be stored air-free
after the evaporation of NH$_{3}$ and handled in a glove-box, as the
products are very air and moisture sensitive, similar to their transition
metal analogues.

The application of large (around 50 \%) molar excess Te instead of excess
acetylide results in a cherry red solution of dissolved species.  In this
case the complete dissolution of the Te powder may take a longer period of
sonication and is believed to produce polytelluride ions with attached
acetylide units.

While Li$_{2}$C$_{2}$ is easy to produce due to the spontaneous
disproportionation of LiC$_{2}$H at room temperature, the production of
heavier bialkali acetylides is increasingly cumbersome, especially for A
$\in$ \{K, Rb, Cs\} as it involves high temperature heating of monoalkali
acetylides mixed with alkali metals in high vacuum
\cite{ruschewitz2003binary,hamberger2012evidence}.  To avoid the direct use
of bialkali acetylides we have developed an alternative synthesis based on
monoalkali acetylides only.  It consists of first reacting the Te powder
with two molar equivalent of monoalkali acetylides according to

\begin{equation} \label{disprop1}
2 AC_{2}H + Te \rightarrow ATeC_{2}H + AC_{2}H
\end{equation} 

that is followed by the disproportionation of the resulting
mixture: 

\begin{equation} \label{disprop2} 
ATeC_{2}H + AC_{2}H \rightarrow A_{2}TeC_{2} + H_{2}C_{2} \uparrow.  
\end{equation} 

The first step of this reaction is familiar from analogous acetylenic
organo-telluride reactions when the H of ATeC$_{2}$H is replaced by an
organic functional group, typically an arene
\cite{radchenko1989acetylenic,gedridge1992preparation,caracelli20104}.  The
second step may largely complete when the solvent NH$_{3}$ is evaporated along
with the acetylene gas, with the possibility that the rest of the acetylene gas is removed 
when the product is ground in a mortar.
A repeated condensation of NH$_{3}$ and sonication
of the mixture, followed by repeated evaporation of NH$_{3}$ helps to better mix
the reaction components and remove as much residual acetylene gas as possible, before grinding the
product in a mortar.
We have not characterized ATeC$_{2}$H, as the above two reactions are part of an
equilibrium and the isolation of ATeC$_{2}$H appeared difficult, if at all
possible.  The identity of the products for both Na$_{2}$TeC$_{2}$ syntheses
are indicated by overlapping powder x-ray diffraction (XRD) spectra.  The
monoalkali acetylide based synthesis of ternary acetylides may be applicable
to the synthesis of a wide variety of ternary acetylides avoiding the
cumbersome production of heavier bialkali acetylides and the use of
high-temperature solid state reactions.

\begin{figure}[tb!]
\resizebox*{6.0in}{!}{\includegraphics{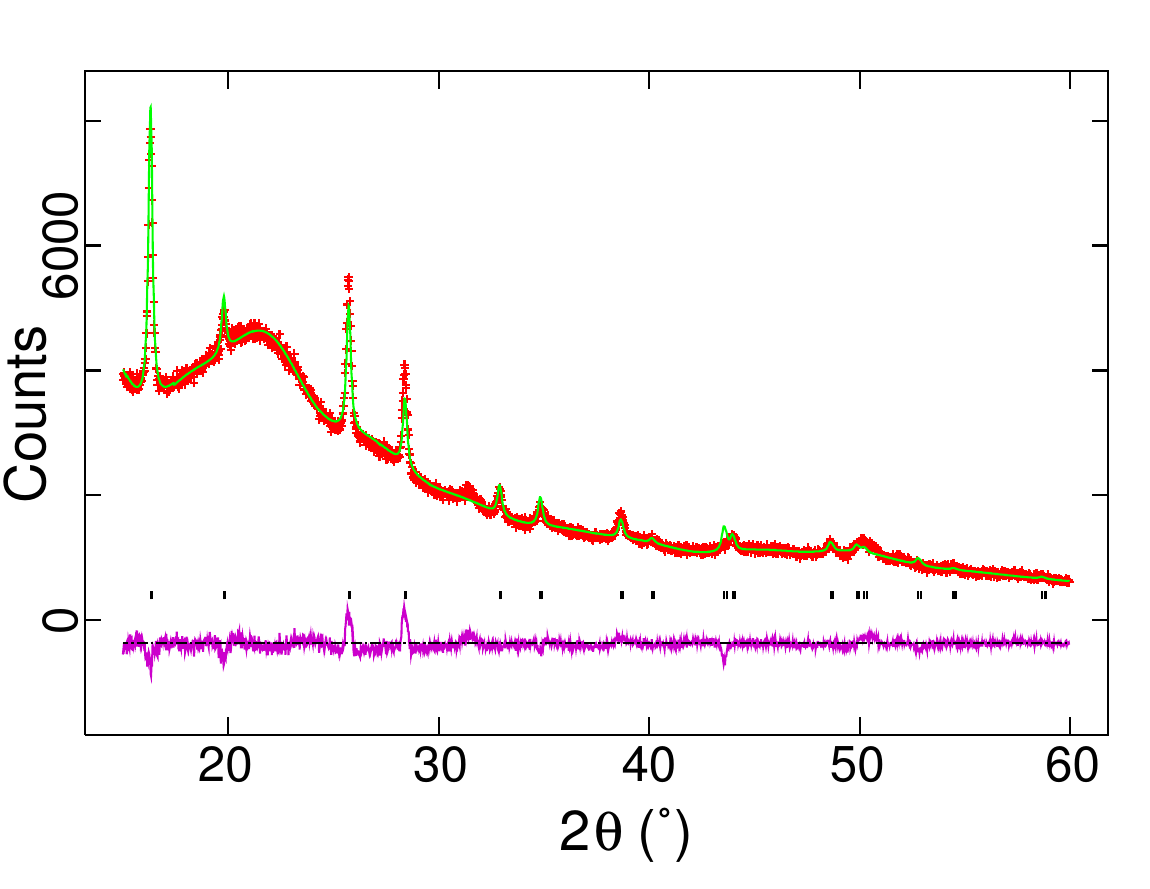}}
\caption{
Powder X-ray diffraction spectrum of Li$_{2}$TeC$_{2}$ (red), the corresponding Rietveld-fit (green)
and the difference of experimental and fitted spectra (purple).
}
\label{XRD-Li2TeC2}
\end{figure}

\begin{figure}[tb!]
\resizebox*{6.0in}{!}{\includegraphics{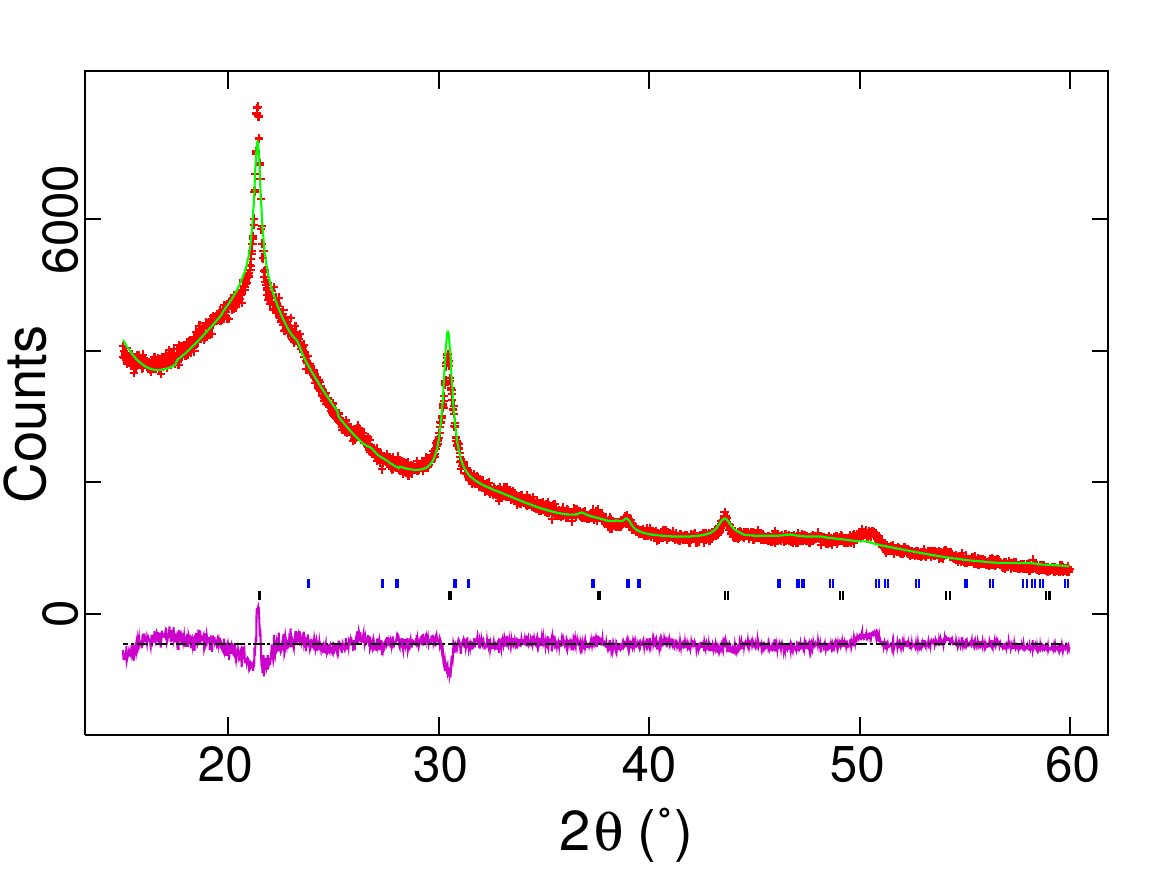}}
\caption{
Powder X-ray diffraction spectrum of Na$_{2}$TeC$_{2}$ (red), the corresponding Rietveld-fit (green)
and the difference of experimental and fitted spectra (purple).
}
\label{XRD-Na2TeC2}
\end{figure}

\begin{figure}[tb!]
\resizebox*{6.0in}{!}{\includegraphics{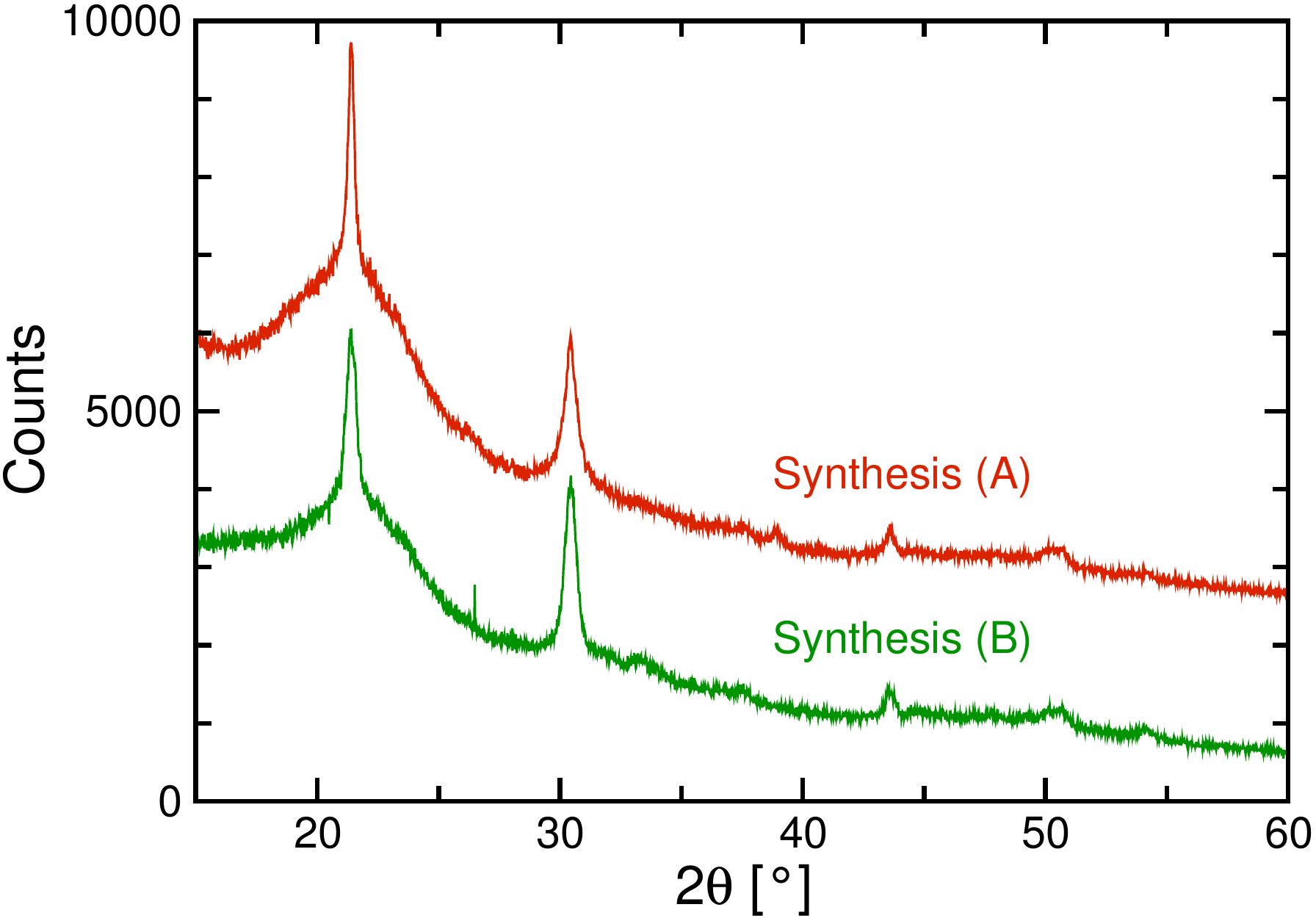}}
\caption{
Powder X-ray diffraction spectra of Na$_{2}$TeC$_{2}$ as obtained by two different synthesis methods.
Synthesis (A): Na$_2$C$_{2}$ + Te $\rightarrow$ Na$_2$TeC$_{2}$. 
Synthesis (B): 2 NaC$_{2}$H + Te $\rightarrow$ Na$_{2}$TeC$_{2}$ + C$_{2}$H$_{2}$. 
}
\label{Comparison-XRD-Na2TeC2}
\end{figure}

\begin{figure}[tb!]
\resizebox*{3.4in}{!}{\includegraphics{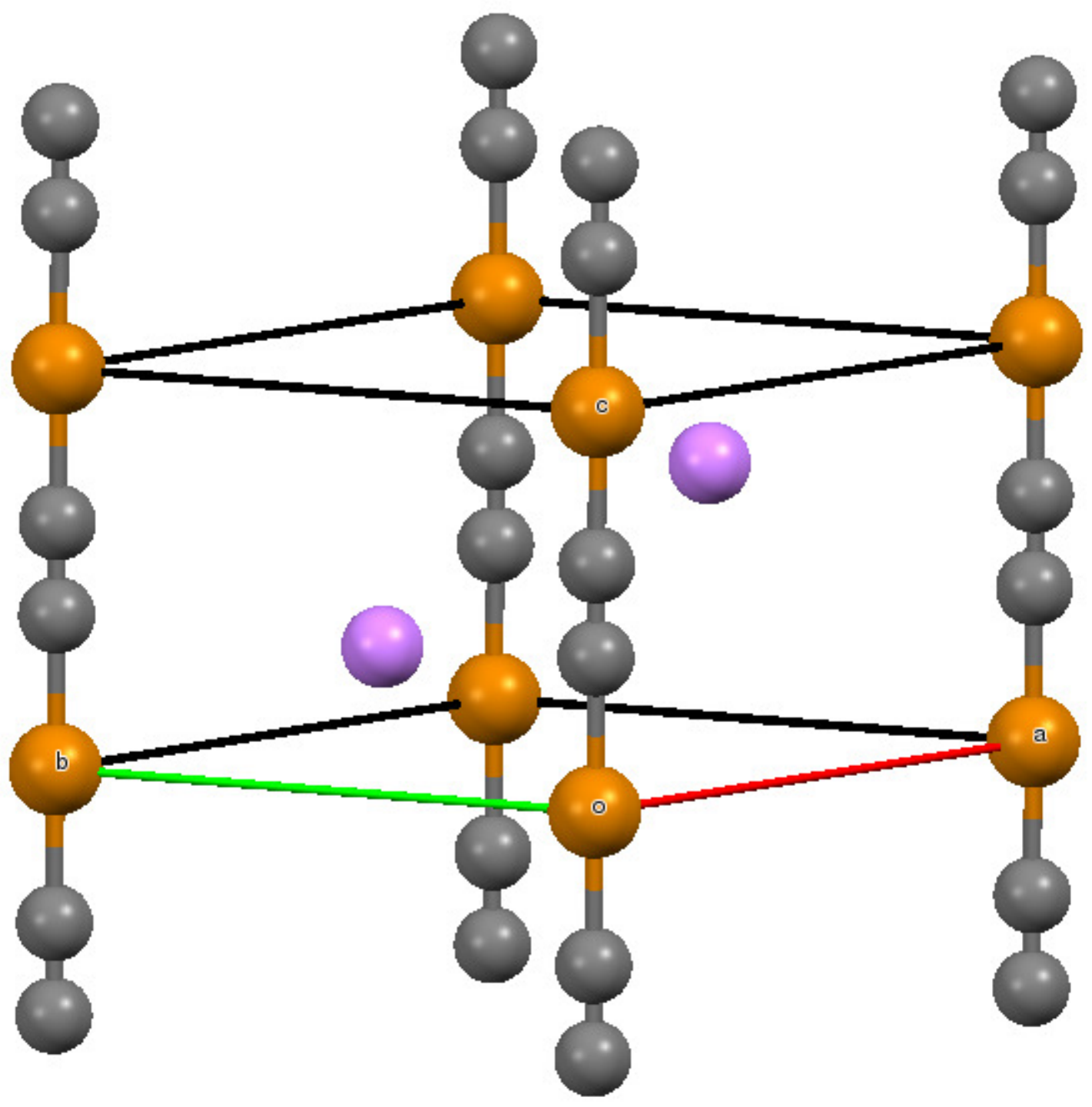}}
\caption{
Stucture of the Li$_{2}$TeC$_{2}$ crystal (space group P$\overline{3}$m1) as determined from the
powder x-ray diffraction data.
Color code: C - gray, Te - bronze, Li - violet.
}
\label{Li2TeC2-Fitted}
\end{figure}

\begin{figure}[tb!]
\resizebox*{3.4in}{!}{\includegraphics{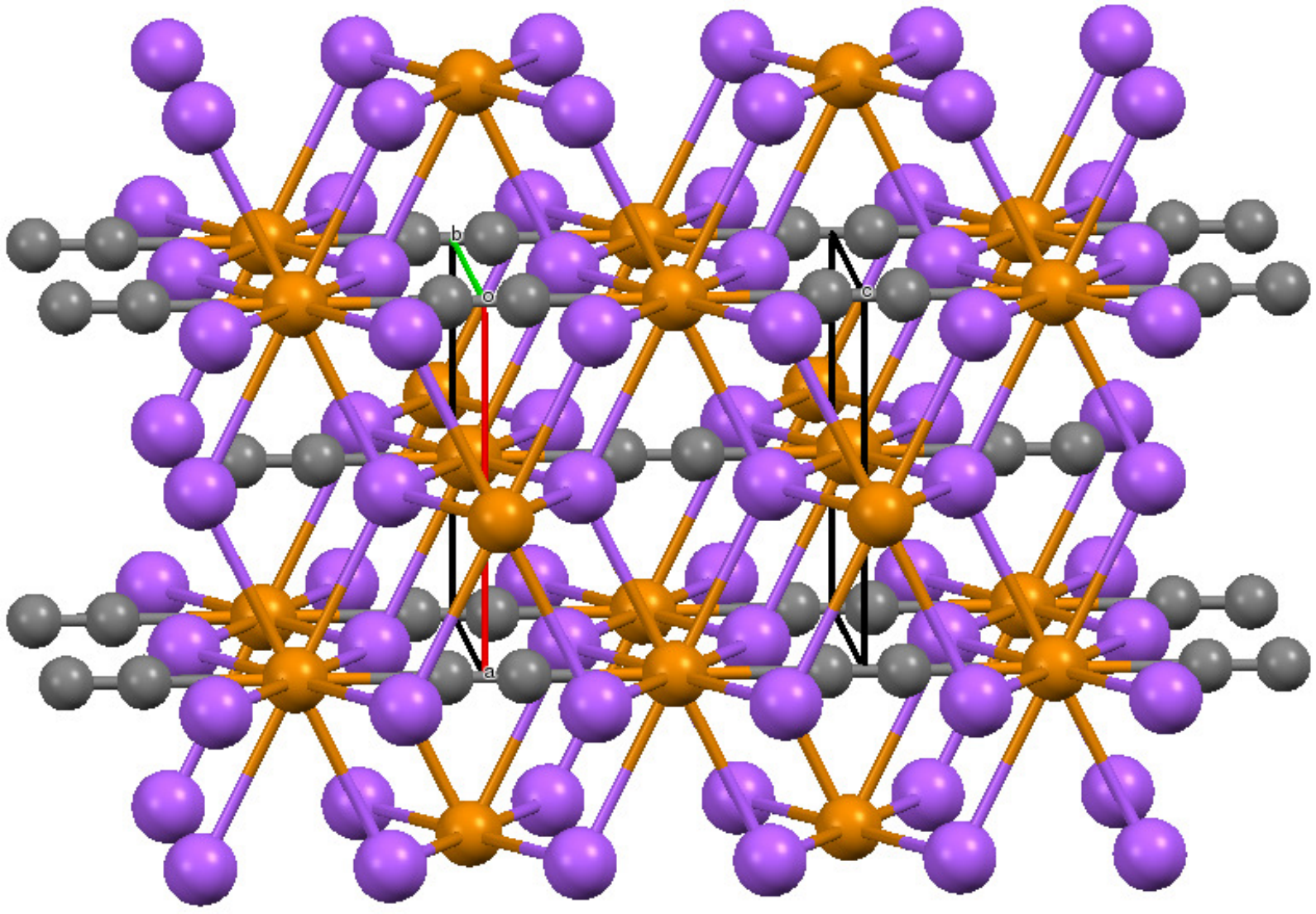}}
\caption{
Stucture of the Na$_{2}$TeC$_{2}$ crystal (space group I4/mmm) as determined from the powder
x-ray diffraction data.
Color code: C - gray, Te - bronze, Na - violet.
}
\label{Na2TeC2-Fitted}
\end{figure}

Powder XRD spectra of Li$_{2}$TeC$_{2}$ and Na$_{2}$TeC$_{2}$ as well as the
comparison of XRD patterns of Na$_{2}$TeC$_{2}$ made by the two different
synthesis routes described above are shown in Figures \ref{XRD-Li2TeC2},
\ref{XRD-Na2TeC2} and \ref{Comparison-XRD-Na2TeC2}, respectively.  The main
geometric parameters of Li$_{2}$TeC$_{2}$ and Na$_{2}$TeC$_{2}$, as obtained
from Rietveld fitting of XRD data, are listed in Table
\ref{Crystalstructures}.  Full details of the fitting are available as
supplementary material \cite{A2TeC2supplemental}.  
Note that XRD data for 2$\Theta$ values smaller than those shown in Figs. \ref{XRD-Li2TeC2}
and \ref{XRD-Na2TeC2} are all due to the sample holder, including a sharper
peak at about 13 and a broader one at about 10 ($^{\rm o}$) 2$\Theta$. 
Figures \ref{Li2TeC2-Fitted} and \ref{Na2TeC2-Fitted} display the fitted structures
of Li$_{2}$TeC$_{2}$ and Na$_{2}$TeC$_{2}$, respectively.  The space group
of Li$_{2}$TeC$_{2}$ is P$\overline{3}$m1, identical with the space group of
all known A$_{2}$MC$_{2}$ (M=Pt,Pd; A=Na,K,Rb,Cs) compounds.  The space
group of Na$_{2}$TeC$_{2}$, however, is I4/mmm, representing a new structure
in ternary acetylides. 
Also note that the fitting of Na$_{2}$TeC$_{2}$ data identified two phases, a major one 
of the actual Na$_{2}$TeC$_{2}$ material with I4/mmm space group and a minor one of the 
unreacted excess Na$_{2}$C$_{2}$ used in the synthesis.

The C$\equiv$C distance in Li$_{2}$TeC$_{2}$ is
1.044 {\AA}, interpreted as a projection of a wobbling C$\equiv$C dumbbell
onto the c axis, therefore it is shorter than an acetylenic bond (C$\equiv$C distance is 1.203 {\AA} in
acetylene gas). The C$\equiv$C
distance in Na$_{2}$TeC$_{2}$, as projected on the a and b axes, is 1.208
{\AA} very close to that in acetylene.  The wobbling motion of the
C$\equiv$C dumbbell and its effect on the projected C$\equiv$C distance has
also been observed in other ternary acetylides, such as K$_{2}$PdC$_{2}$
\cite{billetter2010ternary}.  The short, 1.727 {\AA}, Te-C distance in
Li$_{2}$TeC$_{2}$ should also be interpreted as a projected distance, while
the long, 2.333 {\AA}, Te-C distance in Na$_{2}$TeC$_{2}$ is closer to
expectations based on DFT predictions \cite{terdik2012anomalous}.  The Te-C
distance in bis[(4-methylphenyl) ethynyl] telluride is 2.045 {\AA}
\cite{caracelli20104}, while it is expected to be about 2.4 {\AA} in ternary
acetylides with Te on the basis of DFT calculations, assuming the
P$\overline{3}$m1 space group \cite{terdik2012anomalous}.
The surprisingly large a/c ratio in Li$_{2}$TeC$_{2}$ and the large Li-Te and
Li-C distances appear to correlate with the above mentioned wobbling motion of the C$\equiv$C
unit around the c-axis and may explain also the broad peaks in the XRD spectra.
The broadness of the observed XRD spectral lines may partially 
also be attributed to crystal defects that may occur 
due to the low temperature of the synthesis where the
crystal structure can not anneal well.

In the present phase of our research on ternary acetylides with tellurium,
our only method of characterization was x-ray diffraction of the freshly made samples
shortly after the synthesis, using an air-tight PMMA sample holder. We have also attempted to
obtain synchrotron XRD data of the same samples: unfortunately, the samples partially decomposed
during the storing and handling, this was indicated by significant additional peaks
in the synchrotron XRD spectra, as compared to the locally measured ones. Neutron diffraction
would also be very useful, especially for Li$_{2}$TeC$_{2}$, for a more precise determination
of the Li and C positions, as Te has a much greater x-ray scattering factor. 
The measurement of Raman spectra would be helpful in the characterization of the C$\equiv$C
bond and C$^{13}$ NMR to the types of carbon atoms present. 

The above types of additional characterizations could not be carried out in the present work
primarily for reasons of great air-sensitivity of the product materials. They 
will be subject of forthcoming investigations. 
However, we are still confident that we have identified
the title new compounds solely on the basis of the presented XRD data and through the control
of the stoichiometries of the reactants. The XRD spectra of
related compounds, such as A$_{2}$O, AOH, A$_{2}$Te, A$_{2}$Te$_{2}$, Te, A$_{2}$C$_{2}$
and AC$_{2}$H (A = Li, Na, with the exception of Li$_{2}$Te$_{2}$ that is not known \cite{smith2000syntheses}) 
are so clearly distinct from the observed XRD spectra, that none of them was found to be present
in the syntheses products, except a small amount of the alkali acetylides used in excess as 
reactants. 

We have not taken elemental analysis of the products either, as the purity of the reagents was
confirmed by XRD analysis and the solvent lq-NH$_{3}$ is not expected to react with them at
the low temperatures applied and is unlikely to form stable complexes with Li$^{+}$ or Na$^{+}$ at
room temperature as ammonia complexes of alkali acetylides are known to exist only at the low
temperatures of lq-NH$_{3}$ \cite{hamberger2012evidence} and would decompose when warmed up to room
temperature.
Therefore the total stoichiometry of the products should be the same as
that of the reactants with the exception of the reaction using NaC$_{2}$H as reactant.
In this latter case, however, the identity of the diffraction patterns shown
in Fig. \ref{Comparison-XRD-Na2TeC2} indicates identical composition of the products of the
two different syntheses referenced in the same Figure, which makes the
disproportionation of NaC$_{2}$H to Na$_{2}$C$_{2}$ and H$_{2}$C$_{2}$ as described in Eqs. 
\ref{disprop1} and \ref{disprop2} the only possible explanation of this reaction even without the
explicit detection of H$_{2}$C$_{2}$ in the gases when the lq-NH$_{3}$ solvent is evaporated.

It is well known that Zintl anions (polyanions) of heavier p-field elements,
such as Te, Se, As, Sb, Bi, Pb, etc, are highly soluble in lq-NH$_{3}$ (for
reviews see e.g.  Refs.  \onlinecite{corbett1985polyatomic,marcel2008tin}). 
Such polyanions have also been observed with some transition metal elements,
such as Hg \cite{corbett1985polyatomic}.  This solubility of the Zintl
anions in lq-NH$_{3}$ suggests that more ternary acetylides with metalloid
and transition metal elements may be synthesized in lq-NH$_{3}$ following
analogous procedures to the ones described herein for Te-containing ternary
acetylides.  In principle, these procedures may work also for transition
metals if the transition metal starting material is provided in the form of
sufficiently small polyatomic clusters, for example in the form of
pyrophoric iron, nickel, manganese, etc.
This may provide a practical route for the production
and use of ternary acetylides, for example as anode materials in Li-ion batteries. 

\begin{figure}[tb!]
\resizebox*{4.4in}{!}{\includegraphics{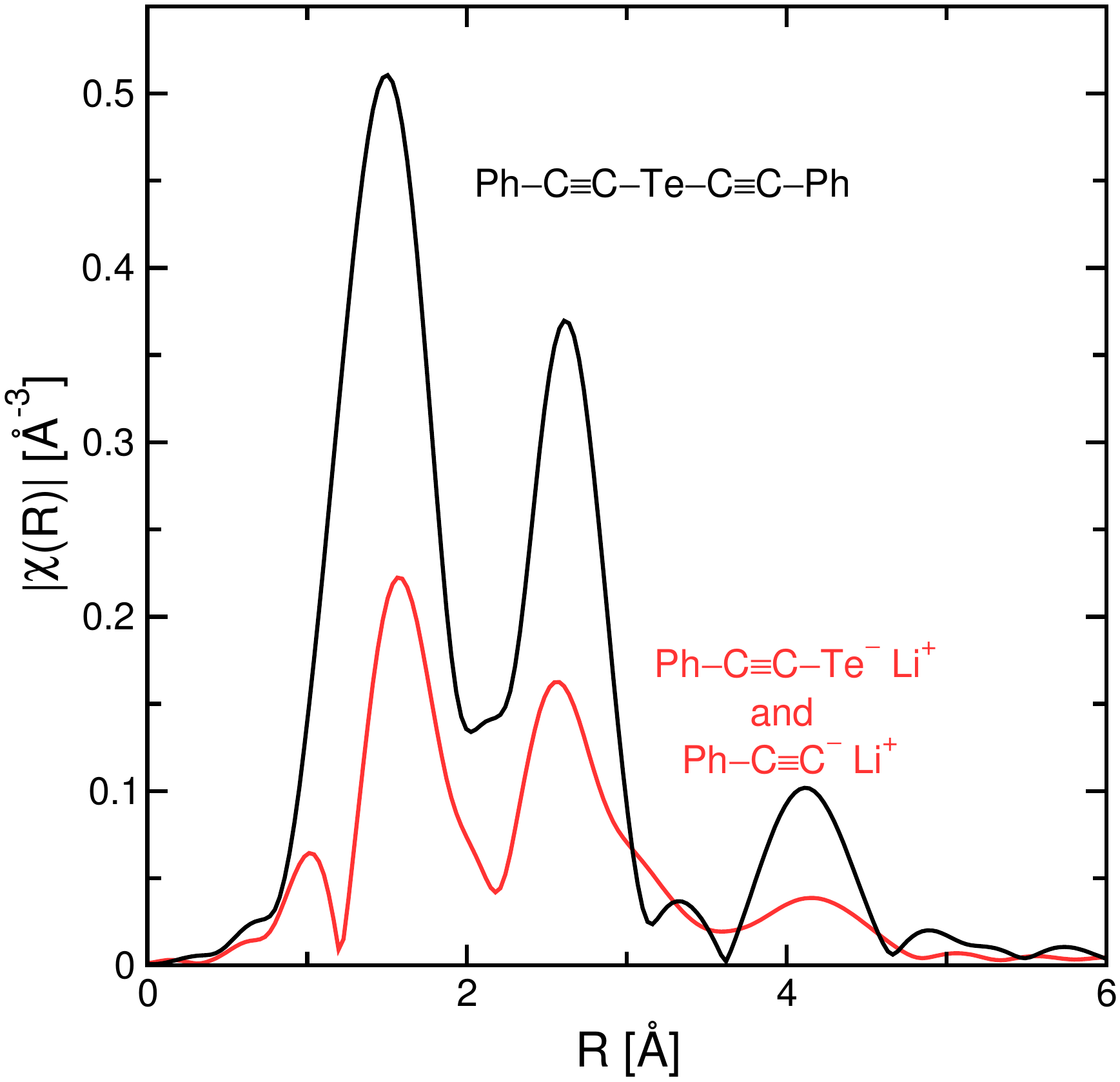}}
\caption{
Comparison of the Te K$_{\alpha}$ EXAFS spectra of
Ph-C$\equiv$C-Te-C$\equiv$C-Ph and its hypothetical double lithiated form. 
The EXAFS intensities of the double lithiated form are approximately 40\% of
that of the unlithiated form, indicating that less than half of the
phenylacetylene groups are bound to Te, i.e.  the double lithiated form is a
mixture of Ph-C$\equiv$C-Te$^{-}$Li$^{+}$ and Ph-C$\equiv$C$^{-}$Li$^{+}$.
}
\label{EXAFScomp}
\end{figure}

In order to test whether the charged bis(ethynyl) tellurides exist in a
molecular form as well, we have synthesized Ph-C$\equiv$C-Te-C$\equiv$C-Ph
following Ref.  \onlinecite{caracelli20104} and its hypothetical double
lithiated form, Li$_{2}$[Ph-C$\equiv$C-Te-C$\equiv$C-Ph], by the direct
reaction of Ph-C$\equiv$C$^{-}$Li$^{+}$ with Te powder in a 2:1 molar ratio
in sonicated dry THF solution.  Figure  \ref{EXAFScomp} shows a comparison of
the Te K-edge EXAFS spectra of the two materials.  The spectrum of
Ph-C$\equiv$C-Te-C$\equiv$C-Ph can be modeled up to $R=3$ \AA\, by nominal
structure which has two C$\equiv$C groups surrounding the Te absorber (see
supplementary material for details, Ref. \onlinecite{A2TeC2supplemental}).  
However the spectrum of
Li$_{2}$[Ph-C$\equiv$C-Te-C$\equiv$C-Ph] shows ca.  0.8 C$\equiv$C groups
surrounding each Te atom on average.  This indicates that the product of the
latter reaction is a less than 1:1 molar mixture of
Ph-C$\equiv$C-Te$^{-}$Li$^{+}$ and unreacted Ph-C$\equiv$C$^{-}$Li$^{+}$,
see Figure  \ref{EXAFScomp}.  This suggests that the negatively charged
-C$\equiv$C-Te-C$\equiv$C- bonding system exists only in the crystals of
A$_{2}$TeC$_{2}$ ternary acetylides and not in molecular forms.

While the ternary acetylides synthesized in the present work may display many interesting
physical properties, such as those predicted in Ref. \onlinecite{terdik2012anomalous} by
Terdik, N{\'e}meth, Harkay, et al., the present work was intended to discuss 
solely the synthesis of these new materials. The examination of their electronic properties,
such as charge distribution, workfunction, photoemissive quantum yield and battery electrode
applications 
should be the subject of separate investigations. We believe that the synthesis methods described
here are significantly new as they allow for a quick solvent-based synthesis of A$_{2}$MC$_{2}$
type ternary acetylides instead of the traditional high temperature solid state synthesis.
The key enabler of the new synthesis methods is the solubility of acetylated polyanionic clusters of 
tellurium atoms (or potentially that of other metalloid or metal elements) in 
lq-NH$_{3}$, described in the present work for the first time in the literature. 
This development was also motivated by 
the recent observation of limited solubility of acetylide ions in lq-NH$_{3}$ by the
Ruschewitz-group \cite{hamberger2012evidence}.


\section{Summary and Conclusions}

We have described two efficient new synthesis methods for the production of
ternary alkali metal tellurium acetylides, Li$_{2}$TeC$_{2}$ and
Na$_{2}$TeC$_{2}$.  The new syntheses methods are based on the reaction of
binary alkali acetylides with tellurium in lq-NH$_{3}$, or on the reaction
of tellurium powder with two equivalent of monoalkali acetylides in
lq-NH$_{3}$ and letting the system disproportionate to the corresponding
bialkali ternary acetylide and acetylene gas. This latter method avoids the
cumbersome synthesis of heavier binary acetylides. The syntheses are
significantly accelerated by ultrasonication.  The colors of the reaction mixture
point toward the existence of polyatomic tellurium anions with attached
acetylene units dissolved in lq-NH$_{3}$ in earlier phases of the reactions. 
The good solubility of these polyanions in lq-NH$_{3}$ and the modest
solubility of binary acetylides allow for fast reactions even at the low
temperatures of the lq-NH$_{3}$ medium, especially when sonication is used,
too.  The ternary acetylides produced are the first examples of ternary
acetylides with metalloid elements, as opposed to the numerous examples of
previously synthesized ternary acetylides with Au, Ag, Cu, Pd and Pt. 
Furthermore, it is expected that these new syntheses may be used analogously
to produce ternary acetylides with other metalloid and transition metal
elements as well, provided that soluble polyatomic anions can be produced in
early phases of the reactions.

\section{Acknowledgements} 

The authors gratefully acknowledge technical help and discussions with Prof. 
A.  Hock, Mr.  M.  Weimer, Mr. M. Foody and other members of the Hock group at IIT, as
well as LiLi Kang and Songyang Han and other members of the Unni group and
technical help from Dr.  B.  Shyam (SLAC/Stanford) and Prof.  J.  Terry
(IIT).  MRCAT operations are supported by the Department of Energy and the
MRCAT member institutions. This research used resources of the Advanced
Photon Source, a U.S. Department of Energy (DOE) Office of Science User
Facility operated for the DOE Office of Science by Argonne National
Laboratory under Contract No. DE-AC02-06CH11357.


\providecommand*\mcitethebibliography{\thebibliography}
\csname @ifundefined\endcsname{endmcitethebibliography}
  {\let\endmcitethebibliography\endthebibliography}{}

\clearpage

\begin{table}[tb!]
\caption{
Selected crystallographic data \cite{A2TeC2supplemental} of A$_{2}$TeC$_{2}$ (A=Li,Na).
The P(C$\equiv$C) and P(Te-C) nearest neighbor distances are projections on the c axis, 
as the C$\equiv$C dumbbell carries out a wobbling
motion, also observed in other ternary acetylides \cite{billetter2010ternary}.
Lengths are given in {\AA}, angles in degrees.
}
\label{Crystalstructures}
\begin{tabular}{ccc}
\hline   
         & Li$_{2}$TeC$_{2}$ & Na$_{2}$TeC$_{2}$  \\
\hline  
Space group    &   P$\overline{3}$m1 &  I4/mmm   \\
 a             &   6.2981(14) & 5.8727(7) \\
 b             &   6.2981(14) & 5.8727(7) \\
 c             &   4.4987(9)  & 5.874(4)  \\
 $\alpha$      &  90.0        & 90.0      \\
 $\beta$       &  90.0        & 90.0      \\
 $\gamma$      & 120.0        & 90.0      \\
 P(Te-C)       &   1.727      &  2.333    \\
 P(C$\equiv$C) &   1.044      &  1.208    \\
 A-A           &   4.131      &  2.936    \\
 A-C           &   3.665      &  3.061    \\
 A-Te          &   3.851      &  3.284    \\
No. of data points & 2226  &  4205     \\
No. of fitting params. &  25  & 19    \\
R$_{p}$    & 0.029 & 0.034 \\
R$_{wp}$   & 0.040 & 0.046 \\
R$_{wp}$   & 0.022 & 0.028 \\
$\chi^{2}$ & 3.312 & 2.269 \\ 
$(\Delta/\sigma)_{max}$ & 3.78 & 1.44 \\

\hline   
\end{tabular}
\end{table}

\clearpage

\clearpage

\clearpage

\clearpage

\clearpage

\clearpage

%
%
%
%
%
%
%
%

\end{document}